\documentclass[twocolumn]{aa}
\usepackage{graphicx}
\begin{document}
   \title{The nearest cool white dwarf ($d\sim4$~pc), 
          the coolest M-type subdwarf (sdM9.5), 
          and other high proper motion discoveries
         \thanks{based on the systematic search in archival data 
         from the SuperCOSMOS Sky Surveys, 2MASS and DENIS, 
         and on spectroscopic observations with the ESO NTT}}

   \titlerunning{The nearest cool white dwarf, the coolest M-type subdwarf, ...}


   \author{R.-D. Scholz
          \inst{1}
          \and
          I. Lehmann
          \inst{2}
          \and
          I. Matute
          \inst{2}
          \and
          H. Zinnecker
          \inst{1}
          }

   \offprints{R.-D. Scholz}

   \institute{Astrophysikalisches Institut Potsdam, An der Sternwarte 16,
              D--14482 Potsdam, Germany\\
              \email{rdscholz@aip.de, hzinnecker@aip.de}
         \and
             Max-Planck-Institut f\"ur extraterrestrische Physik,
             Postfach 1312, D--85741 Garching, Germany\\
             \email{ile@xray.mpe.mpg.de, matute@mpe.mpg.de}
             }

   \date{Received ...; accepted ...}

   \abstract{We report the discovery of seven high proper motion 
             stars with proper motions
             between about 0.7 and 2.2 arcsec/yr, all at relatively low 
             Galactic latitudes ($13^{\circ}<|b|<35^{\circ}$) and located 
             in the southern sky. 
             They were detected in a high proper motion search
             using multi-epoch positions in the optical SuperCOSMOS 
             Sky Surveys and in the near-infrared sky surveys 2MASS and DENIS.
             Classification spectroscopy carried out for six of the 
             objects reveals them to represent three 
             different classes of cool objects in the solar neighbourhood: 
             M dwarfs, M subdwarfs and cool white dwarfs.
             The star with the largest proper motion, SSSPM~J1138--7722, 
             is classified as a very nearby ($d\sim$8~pc) M5.5 dwarf with 
             Galactic thin disk kinematics. A second star with 
             $\sim$2~arcsec/yr proper motion, SSSPM~J1358--3938, is still 
             lacking spectroscopic confirmation but can be classified from
             photometry as thick disk $\sim$M3.5 dwarf. Three objects turn 
             out to be cool subdwarf members of the Galactic thick disk or 
             halo, including the first sdM9.5 object,
             SSSPM~J1013--1356, which represents the currently
             coolest known M subdwarf, another ultra-cool
             subdwarf, SSSPM~J1930--4311, of spectral type sdM7.0 as
             well as an earlier type (sdM1.5) star. The latter, 
             SSSPM~J1530--8146, has an extremely large space velocity
             with clear halo kinematics
             (heliocentric $U,V,W=-534\pm78,-239\pm74,+188\pm23$).  
             Two objects show featureless spectra classifying them as cool 
             white dwarfs with $T_{eff}<4500$~K. One of them, 
             SSSPM~J1549--3544, is an extremely
             nearby ($d\sim4$~pc) thin disk object, the other one,
             SSSPM~J1148--7458, has thick disk kinematics. SSSPM~J1549--3544
             is likely to be the nearest cool white dwarf
             and may be even the nearest isolated white dwarf, i.e. closer 
             than van Maanen~2.

   }

   \maketitle


\section{Introduction}

A large stellar proper motion, of e.g.
$>0.18$~arcsec/yr, which is the lower limit in the New Luyten Two Tenths
(NLTT) catalogue of about 60000 stars (Luyten \cite{luyten7980}), may be an
indication of a small distance and/or an intrinsic high velocity of the star.
Practically, all known stars in the immediate solar neighbourhood ($d<10$~pc)
exhibit proper motions larger than the NLTT limit 
(see http://www.chara.gsu.edu/RECONS/). However,  
a high proper motion (hereafter HPM) 
sample does not reflect the real number densities 
of Galactic thin disk, thick disk and
halo populations in the solar neighbourhood (see e.g. 
Digby et al.~\cite{digby03}) but is strongly biased towards thick disk
and halo members. The larger the lower proper motion limit in such a sample,
the more exotic thick disk and halo objects can be expected.

Therefore, it is not surprising that recent discoveries of new  
extreme HPM objects (with proper motion larger than 
$\sim$2~arcsec/yr; only 77 NLTT stars are above that limit) 
include only a few very nearby ($d<4$~pc) disk M dwarfs 
(Teegarden et al.~\cite{teegarden03}) and T-type  brown dwarfs 
(Scholz et al.~\cite{scholz03}, McCaughrean et al.~\cite{mccaughrean04}). 
On the other hand, there are more 
still relatively nearby ($d < 25$~pc), halo M dwarfs 
(L\'epine et al.~\cite{lepine02})
and L subdwarfs (Burgasser et al.~\cite{burgasser03}) as well as thick disk
or halo cool white dwarfs (Ibata et al.~\cite{ibata00}, 
Scholz et al.~\cite{scholz02a}) being discovered.

Here we present two new discoveries matching the limit of $\sim$2~arcsec/yr
together with five other new HPM stars with 
$0.7 < \mu < 1.7$~arcsec/yr, identified as a result of an ongoing
HPM survey of the southern sky briefly described in 
section~\ref{hpm}. For six of these objects,
which are all located at low Galactic latitudes, as well as for a 
previously known NLTT star apparently moving in front of a dark cloud
close to the direction to the Galactic centre we have carried out optical
classification spectroscopy (sections~\ref{specobs} and~\ref{clspec}). 
For five objects we have measured radial velocities (section~\ref{rvd}).
Section~\ref{dUVWpop} deals with distance estimates on the basis
of the available data, which were combined to determine space
velocities and Galactic population memberships of the objects.
In Section~\ref{concloutl} we draw conclusions and give an outlook. 

   \begin{figure}[ht!]
   \centering
      \caption{(in electronic form only)
         $3\times3$~arcmin$^2$ SSS images (left: $B_J$, centre: $R$, 
         right: $I$) for (from top to bottom): SSSPM~J1013--1356,
         SSSPM~J1138--7722, SSSPM~J1148--7458, SSSPM~J1358--3938, 
         SSSPM~J1530--8146, SSSPM~J1549--3544, T3 and SSSPM~J1930--4311,
         (east is left, north is up).
              }
         \label{8charts}
   \end{figure}
%

 
\section{High proper motion survey and photometry}
\label{hpm}
 
The digitised archival data from the SuperCOSMOS Sky Surveys (SSS)
provide multi-epoch data in three optical ($B_J, R, I$) passbands
(Hambly et al.~\cite{hambly01a,hambly01b,hambly01c}) for the entire
southern hemisphere. These data,
combined with the recently completed near-infrared survey 2MASS
(Two Micron All Sky Survey, Cutri et al.~\cite{cutri03}) are perfectly
suited to the search for hitherto unknown HPM objects
in the southern sky. Different search methods using SSS and 2MASS are
possible.
 
Here we applied essentially a search technique similar to that of
Scholz et al.~(\cite{scholz03}), i.e. starting from SSS $I$-band data and
looking for bright ($I<17$) objects with no counterparts in the $R$- and
$B_J$-bands (with an initial search radius of 3~arcsec) which could be
identified with an other $I$-band measurement on a different plate. The
$6^{\circ}\times6^{\circ}$ UK Schmidt Telescope (UKST) survey plates
overlap by about half a degree on each side so that roughly one third of the
southern sky is covered by overlapping UKST plates. The $I$-band observations
were made starting in the early 80s with most plates taken in the 90s,
whereas the majority of the UKST $B_J$ and ESO $R$ plates (also included in
the SSS) were taken in the 70s and early 80s. The UKST $R$ plates were
typically observed in the late 80s and 90s. Therefore, a search based on
the $I$-band data is sensitive to both HPM and red objects.

The completeness of our HPM survey is difficult to estimate due to
the variation of epoch differences in different bands over the sky.
Hambly et al.~(\cite{hambly04}) have recently published first results
of their most complex HPM search using the complete SSS data base.
From their five new objects with proper motions larger than 1~arcsec/yr
in the declination zone south of $-57.5^{\circ}$, we were able to detect
two, including SSSPM~J1138--7722. On the other hand, there is one object
presented in this paper (SSSPM~J1148--7458), which they did not detect.
Therefore, different search methodologies may complement each other even
if using the same data base.

Altogether seven objects with the largest newly detected proper motions
in the all southern sky HPM survey were selected for this study
taking into account their visibility during the spectroscopic observing
run.
One of the objects (SSSPM~J1358--3938 = 2MASS~13580529--3937545) was first
identified as a potential
HPM candidate among bright 2MASS objects without optical
counterparts in the 2MASS data base.
 
\begin{table*}
 \scriptsize
 \caption[]{Astrometry and photometry from SSS and 2MASS.
}
\label{sss2m}
 \begin{tabular}{lccrrcccrrr}
 \hline
Name & $\alpha, \delta$ & Epoch & $\mu_{\alpha}\cos{\delta}$ & $\mu_{\delta}$ & $B_J$ & $R^i$ & $I^i$ & $J$ & $H$ & $K_s$ \\
SSSPM~J..   & (J2000) & & \multispan{2}{\hfil mas/yr \hfil} & \multispan{3}{\hfil (SSS) \hfil} & \multispan{3}{\hfil (2MASS) \hfil} \\
 \hline
 
1013--1356$^a$& 10 13 07.34 $-$13 56 20.4 & 1999.099 & $  +66 \pm 07$ & $-1026 \pm 04$ & 21.931 & 18.687 & 16.077 & 14.621 & 14.382 & 14.398 \\ 
              &                           &          &                &                &        &(18.752)&(16.406) \\
1138--7722$^b$& 11 38 16.71 $-$77 21 48.4 & 2000.068 & $-2063 \pm 05$ & $ +620 \pm 07$ & 16.280 & 13.897 & 11.290 &  9.399 &  8.890 &  8.521 \\ 
              &                           &          &                &                &        &(13.276)&(11.646) \\
1148--7458$^c$& 11 47 34.44 $-$74 57 59.2 & 2000.068 & $-1078 \pm 07$ & $+1350 \pm 07$ & 18.646 & 17.421 & 16.904 & 15.580 & 15.572 & 15.217 \\ 
              &                           &          &                &                &        &(16.796)&(16.276) \\
1358--3938$^d$& 13 58 05.29 $-$39 37 54.5 & 1999.290 & $+1738 \pm 04$ & $ -889 \pm 03$ & 15.138 & 12.559 & 10.299 &  9.720 &  9.226 &  8.948 \\ 
              &                           &          &                &                &        &(12.724)&(11.218) \\
1530--8146$^e$& 15 30 28.67 $-$81 45 37.5 & 2000.216 & $ -594 \pm 07$ & $ -287 \pm 03$ & 18.511 & 16.636 & 15.715 & 14.154 & 13.601 & 13.404 \\ 
              &                           &          &                &                &        &(16.164)&(15.286) \\
1549--3544$^f$& 15 48 40.23 $-$35 44 25.4 & 1998.625 & $ -591 \pm 08$ & $ -538 \pm 05$ & 15.994 & 14.765 & 14.285 & 12.340 & 11.765 & 11.620 \\ 
              &                           &          &                &                &        &(13.776)&(13.500) \\
1930--4311$^g$& 19 29 40.99 $-$43 10 36.8 & 2000.633 & $  -19 \pm 11$ & $ -865 \pm 08$ & 21.077 & 18.423 & 16.311 & 14.794 & 14.230 & 14.091 \\ 
              &                           &          &                &                &        &(18.411)&(16.359) \\
 \hline
T3$^h$ (NLTT) & 17 15 11.91 $-$27 33 01.1 & 1998.529 & $ -123 \pm 05$ & $ -220 \pm 03$ & 19.904 & 18.408 & 17.210 & 12.514 & 11.318 & 10.827 \\ 
              &                           &          &                &                &        &(17.364)&(15.685) \\
 \hline
 \end{tabular}
Notes:\\
$^a$ -- $\alpha, \delta$ from 2MASS; proper motion from 1 2MASS and 7 SSS positions \\
$^b$ -- $\alpha, \delta$ from 2MASS; proper motion from 1 2MASS and 10 SSS positions \\
$^c$ -- $\alpha, \delta$ from 2MASS; proper motion from 1 2MASS and 7 SSS positions; DENIS $I=16.290, J=15.855$ \\
$^d$ -- $\alpha, \delta$ from 2MASS; proper motion from 1 2MASS, 1 DENIS and 4 SSS positions; DENIS $I=11.161, J=9.651, K=8.929$ \\
$^e$ -- $\alpha, \delta$ from 2MASS; proper motion from 1 2MASS, 4 DENIS and 8 SSS positions; DENIS mean magnitudes: $I=15.243, J=14.156, K=13.397$ \\
$^f$ -- $\alpha, \delta$ from 2MASS; proper motion from 1 2MASS and 6 SSS positions \\
$^g$ -- $\alpha, \delta$ from DENIS; proper motion from 1 2MASS and 4 SSS positions; DENIS $I=16.267, J=14.717, K=13.906$ \\
$^h$ -- $\alpha, \delta$ from 2MASS; proper motion from 1 2MASS, 1 DENIS, 8 SSS and 2 SHS positions; DENIS $I=15.156, J=12.660, K=10.886$ \\
$^i$ -- standard SSS magnitudes are corrected to bring $B_J-R$ and $R-I$ colours to same median across the sky; uncorrected magnitudes given in brackets \\
\end{table*}
 
\normalsize

All new HPM candidates were checked on SSS finding charts 
(see e.g. Figure~\ref{8charts}) in
all passbands with a search radius of up to a few arcminutes.
For a more accurate proper motion determination we used then all identified 
SSS positions on overlapping plates, including those from the SuperCOSMOS
$H_{\alpha}$ survey (SHS) (Parker \& Phillips~\cite{parker98}) in 
the Galactic plane, together with that from 2MASS, and in some cases from 
the second release of DENIS (DEep Near-Infrared Survey, Epchtein et
al.~\cite{epchtein97}) data
(http://vizier.u-strasbg.fr/viz-bin/Cat?B/denis). The astrometric and
photometric data of the extreme HPM objects, which
were selected for the spectroscopic follow-up observations of this study,
are listed in Table~\ref{sss2m}. The SSS magnitudes in the table are
averaged values from several plate measurements. The proper motions
were determined from fitting the available multi-epoch positions. The
errors are relatively small, since for all objects, except for
SSSPM~J1930--4311, there were overlapping plate measurements. 

The standard SSS $R$ and $I$ magnitudes are corrected in such a way that
the SSS $B_J-R$ and $R-I$ colours are anchored at the same median value
across the entire survey (see http://www-wfau.roe.ac.uk/sss/). We also
extracted the uncorrected magnitudes (given in brackets in the second
line after each object in Table~\ref{sss2m}), since the correction may
be wrong for HPM (foreground) stars 
in fields close to the Galactic plane. In some cases, additional
$I$ band magnitudes from DENIS were available (see the notes to 
Table~\ref{sss2m}). At higher Galactic latitudes, we usually 
mentioned good agreement between the (corrected) SSS and DENIS $I$
magnitudes for the majority of HPM objects detected
in our survey. Here, at relatively low Galactic latitudes,
we found more than 1~mag differences for some of
the new HPM stars (see e.g. the large deviation for
SSSPM~J1358--3938). The uncorrected SSS $I$ magnitudes do, however,
agree well with the DENIS $I$ magnitudes. 

One known NLTT star, which showed up with an extreme optical-to-infrared
colour in a cross-correlation of NLTT and 2MASS, was also included in the
target list of our observations. This star, originally
detected by Terzan et al.~(\cite{terzan80}) and listed in Luyten's
NLTT catalogue as T3, is located close to the Galactic centre
($l=357^{\circ},b=+6^{\circ}$) and seems to move in the foreground
of a dark cloud (see images in Figure~\ref{8charts}). We selected this 
object with a proper motion of about 0.25~arcsec/yr, i.e. well above
the NLTT limit of 0.18~arcsec/yr, as a nearby brown dwarf candidate
for our spectroscopic observations. Compared to the large colour indices
$I-J=+4.7$ and $J-K_s=+1.69$, the optical SSS colours were
rather blue: $B_J-R=+1.5$ and $R-I=+1.2$. From the uncorrected magnitudes 
we got the alternative colours: $B_J-R=+2.5, R-I=+1.7$ and $I-J=+3.2$,
which were more consistent with a brown dwarf candidate.

 
\section{Spectroscopy}

\subsection{Observations}
\label{specobs}

Spectroscopic observations of seven HPM stars were
carried out with the 3.5m New Technology Telescope (NTT) at 
ESO La Silla between May 2th and 4th, 2003
using the EMMI instrument equipped with the new 2048 $\times$ 4096 MIT/LL CCDs.
The relatively bright objects were observed at the beginning and end of the
nights as test and supplement targets for the original
programme (ESO N 071.A-0444), dealing with the identification of optically 
faint type-2 quasars candidates selected from XMM-Newton XID fields 
(see Page, Lehmann, Boller et al.~\cite{page04}).
Optical low-resolution spectra were obtained using a 360 line/mm grism
in the RILD mode and a 1~arcsec wide slit. The grism efficiency is 77~\% at 
the blaze wavelength of 4600~\AA. Without an order blocking filter the 
wavelength coverage ranges from 3850-9100~\AA. Second order overlap occurs
beyond 8000~\AA. The spectral dispersion is 
2.3~\AA/pix and the resolution is $\sim5$~\AA. The seeing was between 
0.7~arcsec and 1.4~arcsec. The exposure times ranged from 60~sec for the 
brightest object, SSSPM~J1138--7722, to 300-500~sec for the
other objects.
 
The reduction of the raw spectra was carried out using the standard
routines under the NOAO/onedspec package available in IRAF.  Raw
data were sky-subtracted and corrected for pixel-to-pixel variations by
division with a suitably normalised exposure spectrum of a continuum
lamp. Prior to the extraction and calibration of the spectra, 
the two-dimensional data were corrected
for geometric distorsion, induced by the spectrograph optics, by
fitting the Ne-Ar lamp lines with two-dimensional cubic splines along
the chip of the CCD.  Wavelength calibration was done by
comparison with exposures of Ar and Ne lamps, while relative flux
calibration was carried out by observations of spectrophotometric
standards stars (Hamuy~\cite{hamuy92,hamuy94}). 
The 5577~\AA~sky line was used
to correct for small shifts of the location of the spectra during the
night caused by telescope flexure (generally $< 1$~pixel).

   \begin{figure*}[ht!]
   \centering
   \includegraphics[angle=-90,width=18.0cm]{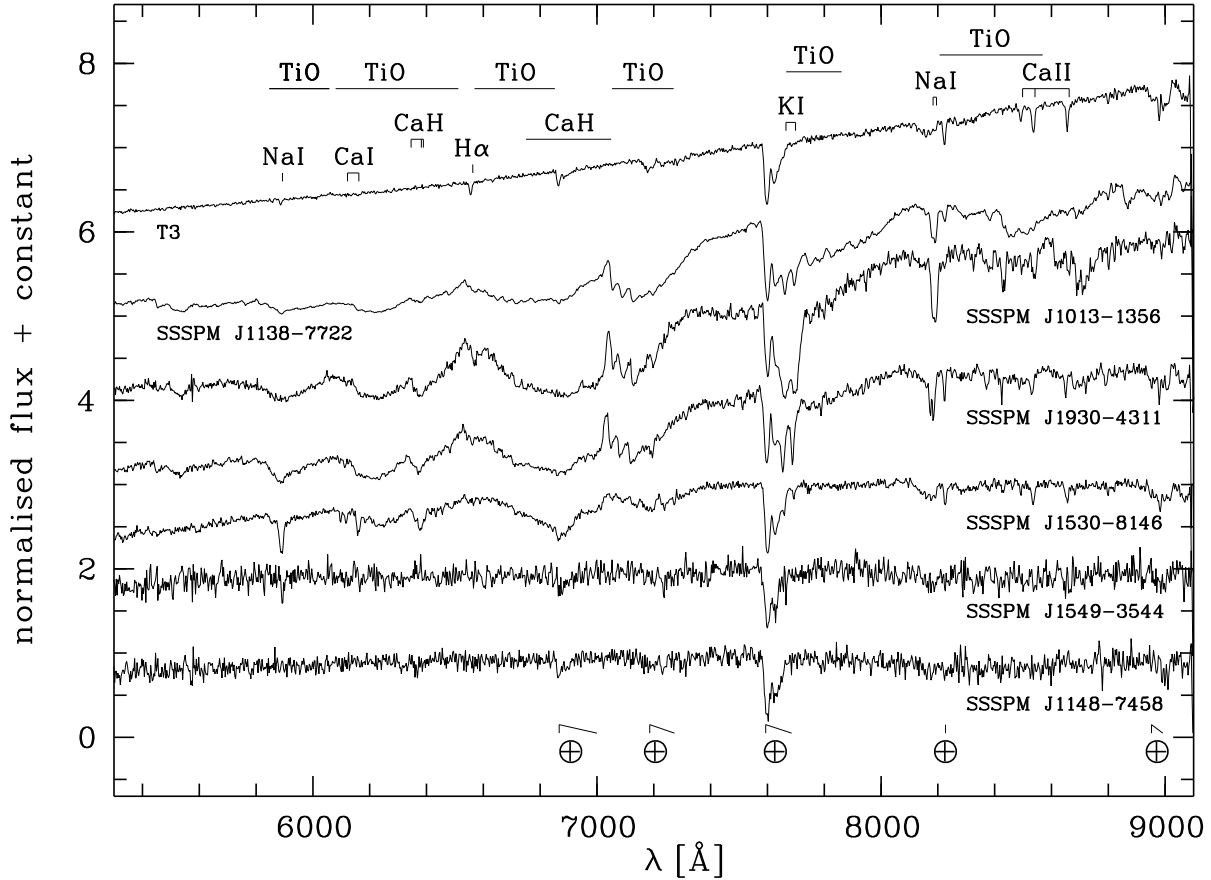}
      \caption{Spectra of seven objects observed with the ESO NTT.
               The spectra are flux calibrated, normalised at 7500~\AA~
               and vertically shifted by integer constants. 
               The positions of the main
               absorption lines are marked at the top, whereas telluric
               absorption bands and lines are marked at the bottom.
              }
         \label{7spec}
   \end{figure*}
%


\subsection{Spectroscopic classification}
\label{clspec}
 
The spectra shown in Figure~\ref{7spec} can be roughly divided into three
different groups: the lower two spectra (of SSSPM~J1148--7458 and
SSSPM~J1549--3544) do not show any features except for telluric absorption
bands; the upper spectrum (of T3) has a very smooth red continuum with
sharp absorption lines resembling reddened earlier-type ($<$K) background 
stars; and the remaining four spectra (of SSSPM~J1138--7722, SSSPM~J1013--1356,
SSSPM~J1930--4311 and SSSPM~J1530--8146) are dominated by broad absorption
bands of TiO and CaH, typical of M dwarfs and subdwarfs.

\subsubsection{Featureless and reddened spectra}
\label{noM}
 
L\'epine et al.~(\cite{lepine03a}) found 11 cool white dwarfs among their 
sample of 104 HPM stars. Half of their spectra are completely
featureless (DC white dwarfs) and half show weak $H_{\alpha}$ absorption
(DA white dwarfs). Our two objects, SSSPM~J1148--7458 and SSSPM~J1549--3544,
can be classified as DC white dwarfs due to the absence of any lines,
although higher signal-to-noise spectra are required to be sure. 
Their effective
temperatures can be estimated to be lower than 4500~K, since both spectra
have a redder continuum than all the cool white dwarfs in L\'epine et
al.~(\cite{lepine03a}). 
We assign a preliminary spectral type of $>$DC11 to each object.

The very red and smooth spectrum on top of Figure~\ref{7spec} looks like
that of a reddened early-type star and was not expected for this relatively
faint HPM object, T3. 
The spectrum was found to be most similar to that of a reddened G-type star, 
since there are no Paschen lines at the red end of the spectrum, typical of 
earlier-type stars, and there are no TiO bands indicative of
later spectral types. Compared to K stars, NaI is only weak and CaI almost 
absent, whereas the $H_{\alpha}$ absorption line is relatively strong
(see Figure~6 in Torres-Dodgen \& Weaver~\cite{torres93}). We can also exclude
the possibility of T3 being a more distant giant, since there is no CN at
7878-8068~\AA~and also no 6497~\AA~blend (BaII, FeI, CaI) left of $H_{\alpha}$, 
which are clearly seen in G5IIIa and G5Ib stars (see Figure~20 in 
Torres-Dodgen \& Weaver~\cite{torres93}). The G2V, G5V and G8V are difficult
to distinguish (as mentioned by Torres-Dodgen \& Weaver~\cite{torres93}) so
that we adopt a spectral type of G5V with an uncertainty of a few subclasses.

\subsubsection{M (sub)dwarf spectra}
\label{Mspec}

A classification scheme for M dwarfs has been developed by 
Reid, Hawley \& Gizis~(\cite{reid95}) by measuring the bandstrengths of TiO 
and CaH. This scheme was later extended to cool subdwarf stars (K and M) by
Gizis~(\cite{gizis97a}) using the spectral indices TiO5, CaH1, CaH2, CaH3
defined by Reid, Hawley \& Gizis~(\cite{reid95}) of observed HPM 
stars mainly from
the Luyten Half second (LHS) catalogue (Luyten~\cite{luyten79}), i.e.
stars with proper motions exceeding 0.5~arcsec/yr. Our NTT spectra have
nearly the same resolution as the spectra used by Reid, 
Hawley \& Gizis~(\cite{reid95}) and Gizis~(\cite{gizis97a}) so that
the classification by the spectral indices can be applied. 

We have computed these indices (see Table~\ref{spkin}) and used them 
to classify the M (sub)dwarfs. Three of them, namely 
SSSPM~J1013--1356, SSSPM~J1930--4311 and SSSPM~J1530--8146, show very
strong CaH (e.g. at $\sim$6350~\AA) in comparison to the TiO band
strengths, giving a strong hint on their probable subdwarf nature. 
Following the three-step classification procedure of Gizis~(\cite{gizis97a}),
we classify these objects as normal subdwarfs (sd) with spectral types
sdM1.5 (SSSPM~J1530--8146), sdM7.0 (SSSPM~J1930--4311) and sdM9.5
(SSSPM~J1013--1356). The latter, extremely late-type M subdwarf,
with its extremely small spectral indices 
(TiO5=0.208, CaH1=0.174, CaH2=0.116, CaH3=0.200),
falls outside the frame of the CaH1/TiO5 and CaH2/TiO5 plots
and is on the edge of the CaH3/TiO5 plot in Figure~1 of 
Gizis~(\cite{gizis97a})!

In order to allow direct comparison with other recently discovered ultra-cool
subdwarfs, we show the CaH2$+$CaH3/TiO5 diagram (Figure~\ref{sd_indices}), 
also used by L\'epine et al.~(\cite{lepine03b,lepine04}), with normal
dwarfs (M), normal subdwarfs (sdM) and extreme subdwarfs (esdM) occupying
different parts of the diagram. Late K (sub)dwarfs are also included
in the upper right part of this diagram. The spectral indices of
SSSPM~J1930--4311 are similar to those of LHS~377, which was the latest-type
(sdM7.0) normal subdwarf in Gizis~(\cite{gizis97a}). Two recently discovered
objects, LSR~J2036+5059 (sdM7.5, L\'epine et al.~\cite{lepine03a}) and
LSR~J1425+7102 (sdM8.0, L\'epine et al.~\cite{lepine03b}) are somewhat cooler
according to their CaH2$+$CaH3 of about 0.5. Our newly discovered
object, SSSPM~J1013--1356 (sdM9.5) is with its extremely small 
CaH2$+$CaH3$\sim$0.3 clearly located below all other objects shown in the
diagram, including the coolest extreme subdwarf, APMPM~J0559--2903 
(esdM7.0, Schweitzer et al.~\cite{schweitzer99}), and is therefore the 
coolest currently known M-type subdwarf. 

Included in Figure~\ref{sd_indices} are all known subdwarfs with 
classification as ''sd'' or ''esd'' and with published
TiO5, CaH2 and CaH3 indices, which we found in the literature 
(Gizis~\cite{gizis97a}, 
Gizis \& Reid~\cite{gizisreid97,gizisreid99}, 
Gizis et al.~\cite{gizis97b},
Schweitzer et al.~\cite{schweitzer99},
Jahrei{\ss} et al.~\cite{jahreiss01},
Cruz \& Reid~\cite{cruz02}
L\'epine et al.~\cite{lepine03a,lepine03b,lepine03c,lepine04}).
Also shown are normal M dwarfs from Gizis~(\cite{gizis97a}), 
Gizis \& Reid~(\cite{gizisreid97}), L\'epine et al.~(\cite{lepine03a}),
and a large number of M dwarfs ($\sim$1850) selected from the sample of
nearly 2000 stars from Reid, 
Hawley \& Gizis~(\cite{reid95}) and Hawley, Gizis \& Reid~(\cite{hawley96})
according to the criteria of Gizis~(\cite{gizis97a}), i.e. his equations (4), 
(5) and (6).

   \begin{figure*}[ht!]
   \centering
   \includegraphics[angle=-90,width=16.0cm]{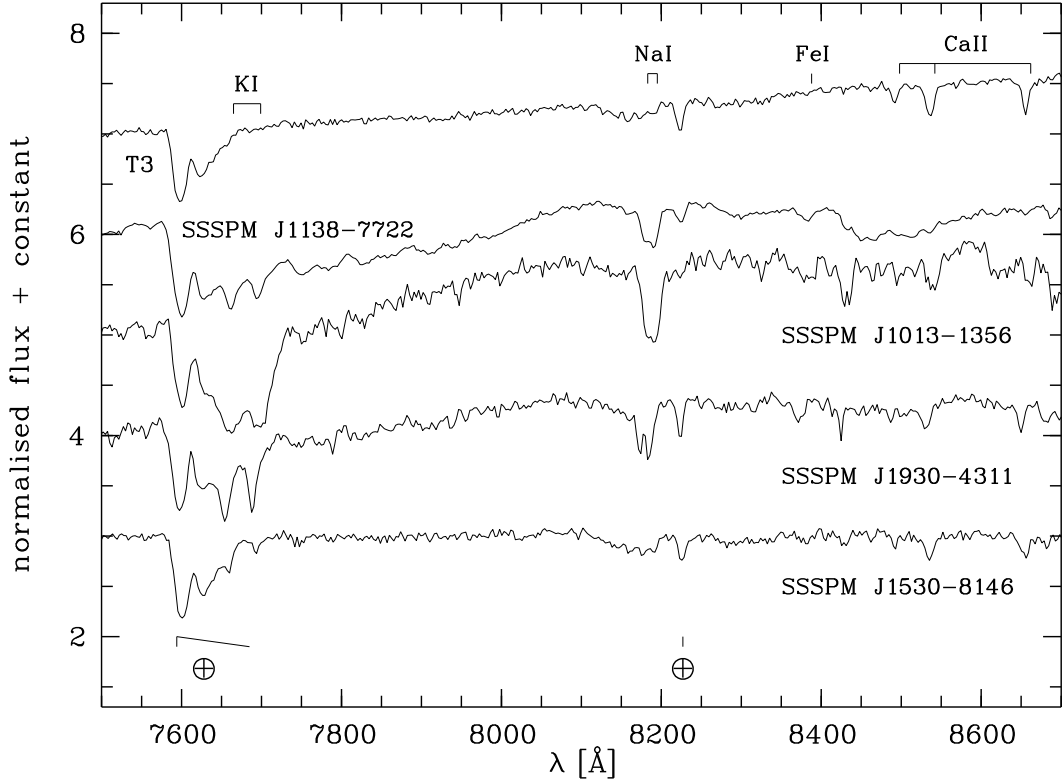}
      \caption{Red spectral region used for radial velocity measurements.
               Shown are the locations of the KI, NaI, FeI and CaII lines 
               which were measured with respect to telluric absorption 
               lines marked at the bottom.
              }
         \label{5spec}
   \end{figure*}
%

SSSPM~J1138--7722, the star with the largest proper motion in our small sample,
is also plotted in Figure~\ref{sd_indices}, where it falls into the region of
normal M dwarfs. A spectral type of M5.5 was assigned to SSSPM~J1138--7722 
according to its TiO5 index and from comparison of the spectrum with those
of known mid-M dwarfs (LHS~168 = M5.0, LHS~546 = M5.5, LHS~1326 = M6.0) also
observed with the ESO NTT with similar resolution 
(Reyl\'{e} et al.~\cite{reyle04}). The direct comparison of the spectra
includes more information than the spectral indices can provide, but
may be affected by flux calibration errors. We have used the wavelength
interval from~5400~\AA~to~8000~\AA, where flux calibration errors as well
as second order spectral overlap are small, and found the spectrum of
SSSPM~J1138--7722 to be in perfect agreement with that of the M5.5 star
LHS~546. The other two comparison star spectra were also confirming
the spectral sequence. Therefore, we adopted the spectral type of M5.5 with
an uncertainty of less than 0.5 spectral subtypes. 

SSSPM~J1013--1356 has a strong red continuum, clearly redder than that
of all other objects in Figure~\ref{7spec}, including that of the only
normal M dwarf in that plot, SSSPM~J1138--7722. On the other hand, we
mention that SSSPM~J1013--1356 is rather blue in the near infrared with
the smallest $J-K\sim0.2$ in Table~\ref{sss2m}.


\begin{table*}
 \scriptsize
 \caption[]{Spectral indices and types, distance estimates and kinematics}
\label{spkin}
 \begin{tabular}{lccccrcrrrrr}
 \hline
Name & TiO5 & CaH1 & CaH2 & CaH3 & ~Sp.Type & $d_{spec}$ & $v_t$  & ${v_r}$  & $U$   & $V$   & $W$   \\
SSSPM~J..   & & & & & & [pc] & [km/s] & [km/s] & [km/s] & [km/s] & [km/s] \\
 \hline
1013--1356 &  0.208 & 0.174 & 0.116 & 0.200 & ~~sdM9.5~       & $50\pm15$  & 245 &  $+6^a$ & $+145\pm63$ & $-140\pm62$ & $-137\pm62$ \\ 
\\
1138--7722 &  0.254 & 0.729 & 0.305 & 0.615 & ~~~~~~M5.5~~    &   8$\pm$1  &  82 & $-50^a$ & $-95\pm16$  & $+2\pm22$  & $+14\pm07$ \\ 
\\
1148--7458 &        &       &       &       & $>$DC11$^c$~~~~ & $36\pm5^d$ & 296 &   $0^g$ & $-185\pm39$ & $-148\pm36$ & $+175\pm35$ \\ 
           &        &       &       &       &                 &            &     & $-50^g$ & $-209\pm39$ & $-106\pm36$ & $+186\pm35$ \\
           &        &       &       &       &                 &            &     & $+50^g$ & $-162\pm39$ & $-191\pm36$ & $+164\pm35$ \\
           &        &       &       &       &                 & $17\pm2^e$ & 140 &   $0^g$ & $-88\pm17$ & $-70\pm23$ & $+83\pm12$ \\
           &        &       &       &       &                 &            &     & $-50^g$ & $-111\pm17$ & $-27\pm23$ & $+94\pm12$ \\
           &        &       &       &       &                 &            &     & $+50^g$ & $-64\pm17$ & $-113\pm23$ & $+71\pm12$ \\
           &        &       &       &       &                 & $18\pm3^f$ & 148 &   $0^g$ & $-93\pm22$ & $-74\pm26$ & $+87\pm18$ \\
           &        &       &       &       &                 &            &     & $-50^g$ & $-116\pm22$ & $-32\pm26$ & $+99\pm18$  \\
           &        &       &       &       &                 &            &     & $+50^g$ & $-69\pm22$ & $-117\pm26$ & $+76\pm18$ \\
\\
1358--3938 &        &       &       &       & ~~~$\sim$M3.5$^b$ & $23^{+11}_{-6}$& 214 &   $0^g$ & $+144\pm53$ & $+78\pm32$ & $-137\pm49$ \\ 
           &        &       &       &       &                 &            &     & $-50^g$ & $+109\pm53$ & $+109\pm32$ & $-155\pm49$ \\
           &        &       &       &       &                 &            &     & $+50^g$ & $+178\pm53$ & $+46\pm32$ & $-119\pm49$ \\
\\
1530--8146 &  0.807 & 0.651 & 0.572 & 0.755 & ~~sdM1.5~       & $185\pm29$ & 581 &$-205^a$ & $-534\pm78$ & $-239\pm74$ & $+188\pm23$ \\ 
\\
1549--3544 &        &       &       &       & $>$DC11$^c$~~~~ & $10\pm2^d$ &  38 &   $0^g$ & $-13\pm23$ & $-36\pm12$ & $-1\pm07$ \\ 
           &        &       &       &       &                 &            &     & $-50^g$ & $-58\pm23$ & $-20\pm12$ & $-14\pm07$ \\
           &        &       &       &       &                 &            &     & $+50^g$ & $+33\pm23$ & $-52\pm12$ & $+12\pm07$ \\
           &        &       &       &       &                 & $3\pm1^e$  &  11 &   $0^g$ & $-4\pm23$ & $-11\pm09$ & $-0\pm07$ \\
           &        &       &       &       &                 &            &     & $-50^g$ & $-49\pm23$ & $+6\pm09$ & $-13\pm07$ \\
           &        &       &       &       &                 &            &     & $+50^g$ & $+42\pm23$ & $-27\pm09$ & $+13\pm07$ \\
           &        &       &       &       &                 & $4\pm1^f$  &  15 &   $0^g$ & $-5\pm23$ & $-14\pm10$ & $-0\pm07$ \\
           &        &       &       &       &                 &            &     & $-50^g$ & $-51\pm23$ & $+2\pm10$ & $-13\pm07$ \\
           &        &       &       &       &                 &            &     & $+50^g$ & $+40\pm23$ & $-31\pm10$ & $+13\pm07$ \\
\\
1930--4311 &  0.288 & 0.354 & 0.210 & 0.387 & ~~sdM7.0~       & $73\pm12$  & 301 &$-262^a$ & $-288\pm25$ & $-270\pm57$ & $+50\pm17$ \\ 
 \hline
\\
T3         &        &       &       &       &  ~~~dG5~~~      & $(70^{+50}_{-20})^h$ & 84 & $-64^a$ & $-66\pm25$ & $-81\pm35$ & $-15\pm05$ \\ 
           &        &       &       &       &                 & $370\pm80^i$ & 444 & $-64^a$ & $-76\pm25 $ & $-438\pm122$ & $-51\pm15$ \\
\\
 \hline
 \end{tabular}
\smallskip

Notes:\\
$^a$ -- Measured radial velocities, errors are about $\pm$25~km/s \\
$^b$ -- Spectral type estimate from colours (see text and Table~\ref{sss2m}) \\
$^c$ -- Preliminary, conservative spectral type estimate \\
$^d$ -- Photometric distance estimate from Equation~(\ref{photd}) using (standard) corrected SSS magnitudes (see Table~\ref{sss2m}) \\
$^e$ -- Photometric distance estimate from Equation~(\ref{photd}) using uncorrected SSS magnitudes (see Table~\ref{sss2m}) \\
$^f$ -- Adopted distance estimate from 2MASS photometry of comparison objects and preferred use of uncorrected SSS magnitudes in Equation~(\ref{photd}) \\
$^g$ -- Assumed radial velocity values with assumed errors $\pm$25~km/s \\
$^h$ -- Assumed distance based on Hipparcos parallax measurements of other G-type main sequence stars with similar proper motions \\
$^i$ -- Assumed distance based on zero extinction in $K_s$ band \\
\end{table*}
 
\normalsize


\subsection{Radial velocity measurements}
\label{rvd}

We were not able to see any lines (e.g. $H_{\alpha}$) in the two cool white 
dwarf spectra with their relatively low signal-to-noise (see Figure~\ref{7spec}).
Therefore, we could not measure their radial velocities. For the other five 
stars, we investigated the red spectral region shown in Figure~\ref{5spec}, 
which includes both strong stellar absorption lines as well as some sharp 
telluric absorption lines, 
for rough radial velocity estimates. The stellar lines used (with different
weights according to the line strengths) were those of potassium (7665~\AA~
and 7699~\AA), sodium (8183~\AA, 8195~\AA), iron (8388~\AA) and
calcium (8498~\AA, 8542~\AA, 8662~\AA). These lines are usually found
in M dwarfs and subdwarfs (see Kirkpatrick et al.~\cite{kirkpatrick91}),
whereas the spectrum of the reddened G-type star, T3, 
shows only a very strong calcium triplet and a weak sodium doublet
in that red spectral region. The telluric H$_2$O line
(8227~\AA) and the sharp absorption feature in the atmospheric A band 
of telluric O$_2$ (at about 7601~\AA), which were also used by 
Scholz et al.~(\cite{scholz02b}), served as reference wavelengths
and for placing the red parts of all seven spectra on the same internal
system.

The accuracy of the radial velocity measurements is dominated by the zero-point
calibration error with the telluric lines and is estimated 
to be about $\pm25$~km/s.
The measured radial velocities have been corrected to a heliocentric 
system and are listed in Table~\ref{spkin}.

 
   \begin{figure}[ht!]
   \centering
   \includegraphics[bbllx=46pt,bblly=81pt,bburx=542pt,bbury=596pt,angle=270,width=8.8cm]{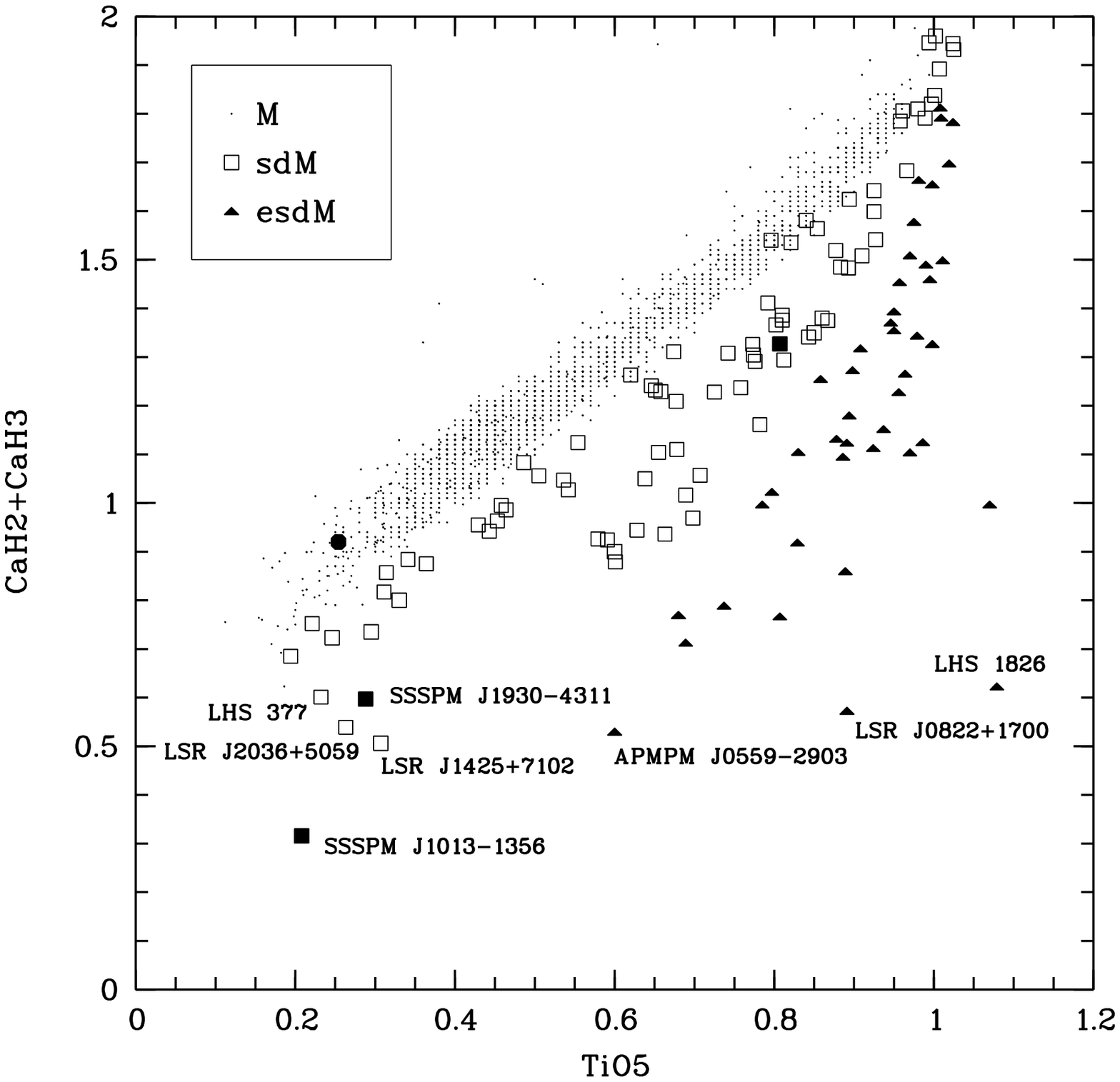}
      \caption{New version of the CaH2$+$CaH3 versus TiO5 diagram, used by
               L\'epine et al.~(\cite{lepine03b,lepine03c}) to discriminate
               between three different object classes (M, sdM, esdM) according
               to the classification scheme of Gizis~(\cite{gizis97a}).
               Included are all known subdwarfs, with the coolest (smallest 
               CaH2$+$CaH3) labeled by their names. The three new subdwarfs
               are marked as filled squares. The new nearby M5.5 dwarf
               is marked by a filled circle. For more details 
               on the samples and individual objects shown in this plot, 
               see text.
              }
         \label{sd_indices}
   \end{figure}
%

 
\section{Distances, space velocities, population memberships}
\label{dUVWpop}

L\'epine et al.~(\cite{lepine03a}) have shown for their
larger HPM sample (including 104 northern stars with
proper motions between 0.5~arcsec/yr and 1.5 arcsec/yr)
the different space velocity distributions of the M dwarfs, subdwarfs and
cool white dwarfs compared to the velocity distributions of Galactic
disk and halo stars from Chiba \& Beers~(\cite{chiba00}).
For the individual objects in our small sample we refer to these data 
(Figures~15 to 18 in L\'epine et al.~\cite{lepine03a}) for comparison. 
Based on the best available measurements (spectroscopic classification, 
2MASS photometry,
proper motions) we have evaluated the distances to the objects, combined
them with the radial velocity measurements (or assumptions)
and computed heliocentric space
velocities (Table~\ref{spkin})
following Johnson \& Soderblom~(\cite{johnson87}). These first
estimates of the kinematics allowed us to identify the new HPM
objects as members
of different Galactic populations (thin disk, thick disk and halo).

 
\subsection{Cool white dwarfs}
\label{cwdUVW}
 
Photometric distances of the newly discovered
cool white dwarfs can be estimated by using the relation
\begin{equation}
\label{photd}
M_{B_J} = 12.73 + 2.58(B_J-R),
\end{equation}
which was derived by Oppenheimer et al.~(\cite{oppenheimer01}). This
relation is valid for SSS magnitudes so that we can use the values given
in Table~\ref{sss2m}. However, we have again the choice between the corrected
(standard) SSS $R$ magnitudes and the uncorrected ones, whereas the $B_J$
magnitudes remain the same. The corresponding distance estimates are shown
in Table~\ref{spkin} (notes~$^d$ and~$^e$), where the formal errors 
correspond to 0.3~mag 
uncertainties in the absolute $B_J$ magnitudes. There is a large difference
between the distances derived from using corrected and uncorrected
SSS $R$ magnitudes. In any case, as can be seen from the 
computed possible space velocities of the cool white dwarfs based on
assumptions on their radial velocities (Table~\ref{spkin}), SSSPM~J1549--3544
is clearly a very nearby ($d=3...10$~pc !) object of the Galactic thin
disk population, whereas SSSPM~J1148--7458 represents more likely
a high velocity object of the Galactic thick disk (if $d=17$~pc)
or halo (if $d=36$~pc).

Not many cool white dwarfs are bright enough to show up in 2MASS
data. This fact alone is already an indication of the probable
proximity of SSSPM~J1549--3544 and SSSPM~J1148--7458. For comparison,
among the 11 cool white dwarfs identified in 
L\'epine et al.~(\cite{lepine03a}) there are only 4 with measured 
2MASS $K_s$ magnitudes, all being comparable to that of SSSPM~J1148--7458
($K_s\sim15.2$) but more than 3~mag fainter than that of SSSPM~J1549--3544
($K_s\sim11.6$). The 2MASS magnitudes are more accurate than 
the photographic SSS magnitudes and can be used for an independent
distance estimate provided there are suitable comparison objects.
SSSPM~J2231--7514 and SSSPM~J2231--7415, a wide pair of cool white dwarfs 
of comparable low effective temperature ($T_{eff}<4000$~K)
discovered recently by Scholz et al.~(\cite{scholz02a}) with an
estimated distance of about 14~pc, have 2MASS magnitudes $J,H,K_s$ of
14.662,14.658,14.436 and 14.858,14.824,14.723, respectively. Using
these values for comparison yields distances of $d=4\pm1$~pc and
$d=20\pm2$~pc, respectively, for SSSPM~J1549--3544 and SSSPM~J1148--7458.
These estimates are in good agreement with those obtained when using the
uncorrected SSS magnitudes in Equation~(\ref{photd}).

Bergeron, Ruiz \& Leggett~(\cite{bergeron97}) list only one
cool white dwarf within 10~pc and with comparable temperature
($T_{eff}=4590$~K), LP~701-29 = LHS~69, with a trigonometric parallax of
$123.7\pm4.3$~mas, for which we find from 2MASS: 
$J,H,K_s$=14.013,13.685,13.546. Using this object as distance calibrator,
we get $d_J=3.7$~pc, $d_H=3.3$~pc, $d_{K_s}=3.3$~pc and $d_J=16.6$~pc,
$d_H=19.3$~pc, $d_{K_s}=17.4$~pc, for SSSPM~J1549--3544 and 
SSSPM~J1148--7458, respectively.

From the above comparisons and preferring the use of uncorrected SSS 
magnitudes in Equation~(\ref{photd}),
we get a good agreement of the distance estimates and adopt $d=4\pm1$~pc
for SSSPM~J1549--3544 and $d=18\pm3$~pc for SSSPM~J1148--7458.
Although trigonometric parallax measurement is required for
confirmation, SSSPM~J1549--3544 is probably the nearest cool white
dwarf and may be even closer than van Maanen~2 = LHS~7, 
the nearest isolated
white dwarf listed in Bergeron, Ruiz \& Leggett~(\cite{bergeron97})
with $T_{eff}=6750$~K and a trigonometric parallax of $232.5\pm1.9$~mas.

 
\subsection{M dwarfs}
\label{dMUVW}
 
In order to estimate the distance to SSSPM~J1138--7722, we have compared 
the observed 2MASS magnitudes with absolute magnitudes of M5.5 dwarfs.
Since there are no M5.5 dwarf data given in Kirkpatrick \& 
McCarthy~(\cite{kirkpatrick94}), we used five single M5.5 dwarfs with measured 
trigonometric parallaxes (LHS~2, LHS~39, LHS~549, LHS~1565, LHS3339) 
with the following mean absolute magnitudes $M_J=9.87, M_H=9.31, M_{Ks}=8.97$ 
(Jahrei{\ss}~\cite{jahreiss04}), which yielded
distances of 8.05~pc, 8.24~pc and 8.13~pc, respectively.
Assuming an uncertainty in the absolute magnitudes of $\pm$0.3~mag, we
adopted a spectroscopic distance of $8\pm1$~pc.

For the second object with a proper motion of about 2~arcsec/yr in our sample,
SSSPM~J1358--3938,
we do not yet have a spectrum, but we can estimate its spectral type from
its $I-J=1.44$ computed from DENIS and 2MASS data (see Table~\ref{sss2m}). 
This $I-J$ colour is typical of M3.5 dwarfs (Kirkpatrick \& 
McCarthy~\cite{kirkpatrick94} give $I-J=1.32$ for M3, $I-J=1.47$ for M3.5
and $I-J=1.52$ for M4).
Assuming a spectral type of M3.5 for SSSPM~J1358--3938, we can derive its
distance from the comparison of the 2MASS magnitudes with corresponding 
absolute magnitudes of M3.5 dwarfs given in Kirkpatrick \& 
McCarthy~(\cite{kirkpatrick94}) to be about $23^{+11}_{-6}$~pc, taking into
account a larger uncertainty in the absolute magnitudes of $\pm$0.8~mag
(corresponding to 0.5 spectral subclasses).

Considering the computed $UVW$ velocities (Table~\ref{spkin}),
SSSPM~J1138--7722 can be identified as a thin disk object.
SSSPM~J1358--3938 seems to be a thick disk object with higher
space velocity according to our distance estimate and independent
of the assumed moderate ($-50...+50$~km/s) radial velocities.

 
\subsection{M subdwarfs}
\label{sdMUVW}
 
For the distance determination of the three subdwarfs we have used 
comparison stars from among the LHS subdwarfs with measured trigonometric
distances listed in Gizis~(\cite{gizis97a}) and extracted their 2MASS
photometry. The following absolute magnitudes $M_J,M_H,M_{K_s}$
were obtained: 7.801,7.316,7.082 (sdM1.5), 10.461,10.000,9.746 (sdM7)
and 11.5,11.0,10.7 (sdM9.5). These values rely on the sdM1.5 stars
LHS~178 and LHS~482, the sdM7.0 star LHS~337 and on extrapolating
the data from the sdM4.5 star LHS3409 via those of LHS~337 (sdM7.0)
to get approximate absolute magnitudes of sdM9.5 objects, respectively.

Assuming for 0.3~mag uncertainties in the absolute magnitudes of sdM1.5
and sdM7.0 and somewhat larger uncertainty (0.4~mag) for sdM9.5 objects,
we got distance estimates of $185\pm29$~pc for SSSPM~J1530--8146 (sdM1.5),
$73\pm12$~pc for SSSPM~J1930--4311 (sdM7.0) and $50\pm15$~pc for
SSSPM~J1013--1356 (sdM9.5).

All three newly discovered cool subdwarfs described in this paper are
remarkable in different aspects. SSSPM~J1013--1356 is clearly the coolest
M-type subdwarf known so far, as can be seen from its separate location
in Figure~\ref{sd_indices}, well below all other objects.  
The other late-type (sdM7.0) subdwarf, SSSPM~J1930--4311, was found to
be member of an extremely wide common proper motion pair, which will be
further investigated in a separate paper (Scholz et al., in preparation).
The third subdwarf, SSSPM~J1530--8146 (sdM1.5), is not as cool as the other 
two but has the largest space motion of all three (see Table~\ref{spkin})
and thus is clearly a Galactic halo representative, whereas the other two
may be members of the thick disk or halo component.


\section{Conclusions and outlook}
\label{concloutl}

Combining optical (SSS) and near-infrared (2MASS, DENIS) archival data
has allowed us to discover some extreme HPM stars close
to the Galactic plane. From the analysis of low-resolution spectra and
from the available photometry/astrometry we conclude that among these 
objects there are:

   \begin{itemize}
      \item two featureless 
            cool white dwarfs, including the probably closest
            ($d\sim4$~pc) white dwarf with $T_{eff}<4500$~K,
            which may also be the closest isolated white dwarf
      \item two nearby M dwarfs with proper motions of $\sim$2~arcsec/yr, 
            one M5.5 dwarf at 8~pc and
            one probably within 25~pc (still 
            to be confirmed spectroscopically)
      \item three M subdwarfs, including the currently coolest known
            M subdwarf, for which we assigned a spectral type of sdM9.5
   \end{itemize}

The NLTT star T3 was rejected as a foreground brown dwarf
candidate but instead classified as a reddened G-type star,
possibly associated with a dark cloud. It deserves further
attention, since the dark cloud may be located relatively nearby.
Alternatively, the star could move with a large space velocity.
Special high resolution spectroscopy in the region
of interstellar absorption features (see e.g. Hearty et al.~\cite{hearty04})
could help to estimate the distance to the cloud and to separate
fore- and background objects.

All the new HPM stars are certainly 
worth to be investigated further by
more detailed follow-up observations. In particular, higher-resolution
spectroscopic observations will yield more accurate radial
velocities for the M (sub)dwarfs and could also help to confirm the
kinematics and hence the Galactic halo or disk membership
of the cool white dwarfs, if any lines can be detected in their
spectra. 

The nearby cool white dwarfs, SSSPM~J1549--3544 and
SSSPM~J1148--7458, and M dwarf (candidates), SSSPM~J1138--7722 (and
SSSPM~J1358--3938), obviously present promising targets for
ongoing trigonometric parallax programmes aimed at filling the
census of the nearby stars sample and at the search of low-mass
companions. The nearest objects are also excellent targets for
extra-solar planet search programmes.

Trigonometric distance estimates, although
more difficult to obtain, would also be important for 
the cool subdwarfs, again in order to constrain their kinematics,
in particular to check the extremely large space velocity of
the sdM1.5 object SSSPM~J1530--8146, which is dominated by the
tangential velocity, derived from the accurate proper motion and
the rather uncertain distance estimate.

The sdM9.5 object, SSSPM~J1013--1356, is of special interest,
since it could belong to the still poorly investigated class of
substellar subdwarfs. Until now, there
is only one clearly sub-stellar L-type subdwarf, discovered by Burgasser
et al.~(\cite{burgasser03}), for which however the classification scheme
of Gizis~(\cite{gizis97a}) could not be applied. Another object,
LSR~1610--0040, according to its spectral indices is only a sdM6.0, 
but was classified by L\'epine, Rich \& Shara (\cite{lepine03c})
as an early L subdwarf on the basis of its very red continuum
and two prominent atomic lines of Rb~I, normally seen only in L dwarfs.
One has to say that the classification scheme of very late-type subdwarfs
(including an extension to sdL) has not yet been developed. 

Among the coolest sdM and esdM objects (marked on Figure~\ref{sd_indices})
there are only two LHS stars but six recently discovered new HPM
objects. This underlines the importance and the potential
of new HPM surveys for completing our knowledge on 
ultra-cool subdwarfs.


\begin{acknowledgements}
This research is based on data from the SuperCOSMOS Sky Surveys
at the Wide-Field Astronomy Unit of the Institute for Astronomy, University
of Edinburgh. We have also used
data products from the Two Micron All Sky Survey, which is a joint project of
the University of Massachusetts and the Infrared Processing and Analysis
Center/California Institute of Technology, funded by the National Aeronautics
and Space Administration and the National Science Foundation, and from
the DEep Near-Infrared Survey (DENIS).
The spectroscopic observations were carried out with the ESO NTT
(as backup programme of ESO N 071.A-0444). 
We thank the anonymous referee
for useful comments, which helped to condense the paper.
\end{acknowledgements}


\end{document}